\documentstyle[emulateapj,psfig]{article}

\begin{document}

\newcommand{\be}{\begin{equation}}
\newcommand{\ee}{\end{equation}}
\newcommand{\nn}{\mbox{} \nonumber \\ \mbox{} }
\newcommand{\ba}{\begin{eqnarray}}
\newcommand{\ea}{\end{eqnarray}}
\newcommand{\Alfven}{ Alfv\'{e}n }
\newcommand{\om}{\omega}

\title{ Radio Emission from Magnetars}
\author{Maxim Lyutikov}
\affil{Physics Department, McGill University, 3600 rue University
Montreal, QC,  Canada H3A 2T8, \\
 Massachusetts Institute of Technology,
77 Massachusetts Avenue, Cambridge, MA 02139, \\
 CITA National Fellow}

\date{Received   / Accepted  }

\begin{abstract}
We discuss properties of the expected
 radio emission from Soft Gamma-ray Repeaters (SGRs)
 during their bursting activity
in the framework of the model of Thompson, Lyutikov and Kulkarni (2002),
in which
 the high energy emission is powered by the
dissipation of super-strong magnetic fields in the magnetospheres 
through 
reconnection-type  events.  Drawing on analogies with Solar flares
we predict  that coherent  radio emission  resembling
 solar type-III radio bursts may be emitted in SGRs  during  X-ray bursts.
The radio emission should have  correlated pulse profiles
with X-rays, narrow-band-type radio spectrum with
$\Delta \nu \leq \nu$  with the typical frequency $ \nu \geq 1 $ GHz,
and, possibly, a drifting central frequency.
We encourage sensitive radio observations of SGRs during  the
bursting activity.
\end{abstract}

\section{Introduction}

Soft Gamma-ray Repeaters (SGRs) and Anomalous X-ray pulsars (AXPs) 
have been identified as isolated magnetized  neutron stars -- 
magnetars (for
recent reviews see,e.g.,  Thompson 2001). Both SGRs and AXPs
have spin periods in the range $P = 6 - 12$ s,
characteristic ages $P/2\dot P = 3\times 10^3 - 4\times 10^5$ yr, and
X-ray luminosities $L_X = 3\times 10^{34} - 10^{36}$ erg s$^{-1}$.
SGRs  are characterized primarily by
occasional repeating bursts of soft gamma rays, as well as by rare
giant gamma ray outbursts that are at least two orders of magnitude
higher in fluence than the smaller events (two have been detected so
far).  The more common small
amplitude bursts have durations of less than $\sim$1~s, have rise times of
typically a few tens of msec, and have fluences that are roughly
correlated with duration (G\"og\"us et al. 2001). In
quiecence, SGRs display X-ray pulsations with periods in the range
5-8~s, spin-down rates in the range $10^{-11}-10^{-12}$, and X-ray
emission, most prominent below $\sim$10~keV, that is well-described by
a power law with photon index $\sim$2. 

Recently Gavriil et al. (2002) have reported an observation
of bursts from AXP. If confirmed, this would establish  a close 
relation between the AXPs and SGRs; similarity between the burst
properties of both classes argue in favour of the same  mechanism of the
X-ray burst production. Below we concentrate on  a better studied
case of SGR bursts, but most of the arguments given may be
applied to AXP bursts as well. 

The radio counterpart status of SGRs has been controversial.  Shitov
(2000) reported the detection of pulsed emission from
SGR~1900+14 at 111 MHz with the Pushchino Radio Observatory.
However, using Arecibo in 1998, Lorimer \& Xilouris (2000)
 observed SGR 1900+14 yet detected no such radio pulsations.
Indeed, no radio pulsations have been detected from any of the SGRs.
This is somewhat surprising since recent radio surveys have discovered
pulsars with polar magnetic
 fields approaching $10^{14}\,$G (Camilo  et al.
2000),
continuous with the lower range of fields deduced from AXP spin-down.

Recently Thompson, Lyutikov and Kulkarni (2002) have proposed a model of the 
SGRs based on the dissipation of the internal super-strong magnetic field,
 generated by
a hydromagnetic dynamo as the star is born, by external currents flowing 
in the magnetosphere. They argued that the currents supporting the
strongly twisted field inside the neutron stars are gradually transported
into  the external magnetosphere, where they can be efficiently dissipated.
The rate with which the currents are transported into the
magnetosphere depends 
 on the tensile strength of the neutron star crust and the
strength of the non-potential (current-carrying) magnetic fields. 
Two regimes are possible: for plastic-type deformations of the crust the 
twist is implanted at a more or less constant rate, while for 
fracturing-type deformations the twist is implanted in sudden events.
Overall, the behavior of the magnetic field resembles that of the Sun,
as the current is transported from the matter-dominated star into the
magnetically dominated corona. 
The parallels between the  dynamics of the solar and magnetar field loops 
extends even further: in both cases the footpoints are believed to be moved
by the torques acted upon them by the twisted magnetic fields
(in addition, on the Sun, some footpoints are moved around
 by the  convective motions).

With  reservation for our understanding of 
reconnection and particle acceleration, we 
propose here that  
the bursting activity of  AXPs and SGRs is due to  the
reconnection-type events in which 
magnetic energy stored in the non-potential magnetic field
is released in  the magnetosphere.
Pushing the analogies with the Sun even further, 
 we argue that both the  persistent emission and the 
flares, including giant flares,
 may result from discreet energy releasing events,
in which the external field relaxes to a lower energy state with a different
field line 
topology. 
This requires that the external magnetic shear build up gradually, and
that the outer crust of the neutron star is deformed plastically by internal
magnetic stresses.
The energy stored in the external
twist then does not  need to
 be limited by the tensile strength of the crust, but instead
by the total external magnetic field energy.

Any suggestion of the  importance of 
reconnection in  astrophysical sources may only be
based on the empirical relations obtained from Solar observations.
The Solar magnetosphere structure and temporal behavior
 is extremely complicated, as  is beautifully illustrated by the
latest images from the  SOHO and TRACE satellites \footnote
{
$\mbox{ http://sohowww.nascom.nasa.gov/   }$}
\footnote{
$\mbox{http://vestige.lmsal.com/TRACE/ }$
}
. 
It seems almost impossible
to predict the behavior of the magnetosphere. Yet, there
seem to be fairly general   scaling laws, which extend from the
 smallest scales
of the solar flares to magnetically active stars (e.g. T Tauri),
 that relate, for example, the 
magnetospheric activity as observed  through high energy
emission   to the total magnetic flux and radio emission.

The two such correlations that we will rely on  are (i) the
 linear  dependence of the X-ray luminosity  on the
magnetic flux (e.g., Johns-Krull et al. 2000),
 which shows over 10 decades in X-ray
and magnetic fluxes; and (ii)
 a strong correlation between the radio
activity and  the high energy activity, also extending from solar flares to
stars. On the Sun the 
radio intensity of large solar flares, when observed, is
linearly  proportional
to the X-ray flux (Sakurai 1974).
Magnetospheric behavior 100 times more active both in terms of largest flares
and flare frequency has been observed.

The direct consequence of the reconnection is the generation
of radio emission, which always accompanies solar X-ray flares.
The natural prediction is then that the 
 radio emission should be observed from SGRs
during  bursts. 
Below we concentrate on SGRs, keeping in mind that burst may have already
been observed from AXPs (Gavriil et al. 2002). 
Here we discuss the expected  properties of the 
radio emission of the SGRs, offer the best strategy for their detection
and discuss possible effects that may prevent the detection
of  radio emission from SGRs.

\section{Solar flares}

Energy dissipation in solar flares is a generically non-linear phenomena:
 explosive-type instabilities  initially grow exponentially (and thus 
are often called ``avalanches'')
 and saturate after a few e-folding growth times,
 after all locally available free energy has been exhausted (Priest and 
Forbes 2002). 
The mechanism responsible for impulsive reorganization is not established, 
yet it seems that the
dissipation takes place in  spatially separated, unresolved complex
structures, including small scale structures.
X-ray and optical observations of the Sun show fine scale strands down to the
instrument resolution ($\sim 30$ km -- micro and nano-flares),
 implying that the elementary heating 
processes are still unresolved.
Thus, the
Solar corona is active on all scales, from  the 
solar radius, to granular convection
and subarcsecond flux tubes.
 It is still not clear whether the 
continual dissipation of the small-scale current sheets dominates the
heating of the  the
closed field line regions of the corona, or whether the
 dissipation is dominated by the large scale  current sheets, which involve
a global rearrangement of the corona.

Statistical studies of solar energy release events, 
e.g. the distribution of event number versus energy 
content as observed at hard X-rays, have led to a 
description in terms of avalanches in a corona which
 has stored energy and is in a state of self-organized criticality
 (Lu \& Hamilton 1991).
\footnote{Alternative interpretations include the early model
of  Rosner \& Vaiana (1978)  and Aschwanden et al. 1998}
The power-law distribution $N(E) \sim E^{\alpha}$
 naturally follows from such a  model since the
 system
 under consideration has no characteristic spatial scale 
above the elementary scale of the smallest avalanche 
(the smallest energy release event), 
up to the system size, the size of active regions.

\section{Persistent and burst emission from reconnection}

A number of facts points to the  magnetospheric origin of SGR
burst (and possibly persistent) emission.
The short rise time of SGR bursts may be explained in the magnetar model
only if originating in the magnetosphere.
In  case  of persistent emission,
 the initial energy release may  also happen
in the  magnetosphere due to unresolved small scale events. 
Later, the energetic particles will heat the crust that
would produce the thermal emission.

The studies of the statistics of the SGR bursts from SGR 1900+14 
(G\"og\"us et al. 1999) have found a dependence  similar to that of  solar flares
 of the number
of  bursts on their fluences with a power-law index $1.66$ 
over 4 orders of magnitude.
The distribution of time intervals between successive
 bursts from SGR 1900+14 is  also consistent with a log-normal distribution.

Another type of correlation  expected in the reconnection  model 
is the correlation between the burst duration and total release energy. 
This is a natural correlation since larger bursts are required to tap 
into  larger volumes of the energy reservoir.
Such a correlation indeed was seen in the SGR bursts by 
G\"og\"us et al. (1999)
who concluded that
``in all [...] statistical properties, SGR bursts resemble earthquakes
 and solar flares more closely than they resemble 
any known accretion-powered or nuclear-powered phenomena''. 

Other circumstantial evidence favoring  magnetospheric emission
includes
(i) SGR bursts come at random phases in the pulse profile Palmer 
(2000) - this is naturally
explained if (even  only one!) 
 emission cite is located high in the magnetosphere, so that we see all
the bursts 
 (if the bursts were associated with a particular active region
on the surface of the neutron star,
 one would expect a correlation with a phase);
(ii) pulsed fraction increases in the tails of the strong bursts, keeping
the  pulse profile similar  to the persistent emission (Woods 2002) - 
this is easier explained if the energy release processes occurring  high in 
the magnetosphere after the giant burst 
 are connected to the same hot spot on the surface of the NS as the field
which are active during the quiescent phase.
(iii) weak black-body component in the tail of the strong bursts
is more consistent with the magnetospheric emission.
(iv)  smaller fluence SGR events, have
harder spectra than the more intense ones (G\"og\"us 2002) (this is 
 also true for the spikes of multi-structured bursts);
 this is consistent with
short events being due to reconnection,  while longer events have a
large contribution from the surface, heated by the precipitating particles.

\section{Radio emission from flares}

Solar flares release magnetic energy in three equally important
channels: thermal heat, bulk motion  of plasma and energetic supra-thermal
(and/or accelerated)
particles. Solar flares are often accompanied by radio bursts, most often
by  what is called Type-III bursts (Bastian et al. 1998).
Type-III radio bursts are signatures of energetic electrons
 generated during solar flares, traveling along the 
magnetic coronal field lines
\footnote{
Particle acceleration is, in fact, not necessary for the Type-III bursts.
It may
be generated either  by the thermal component 
of the plasma heated by the reconnection  - as the faster particles leave the
reconnection region  conditions for the bump-in-tail
instability develop not far from the reconnection region) or  
 by the fast supra-thermal  particles accelerated by the DC
electric fields in the reconnection region. 
}.
As a result, electrostatic plasma  turbulence develops. 
Electromagnetic radio emission is generated in the collision of two
plasma waves. The resulting emission is a narrow-band emission above  the
second plasma harmonic $ \om\sim 2 \om_p$.
 We propose that similar coherent emission may
be generated in SGRs.
Since the radio emission is generated by the electrons accelerated at the
reconnection cite, we predict that 
if the radio emission is detected during the bursting phases of SGRs,
 its intensity and profile  will be strongly correlated 
with the X-ray bursts.

Generally, two types of particles acceleration have been proposed in
the context of solar flares: (a) acceleration by parallel electric fields and 
(b) stochastic  drift  acceleration by perpendicular electric fields. 
DC acceleration is expected to work as well  in SGRs, while the stochastic 
one
is likely to be suppressed by the super-strong magnetic fields.
We do not
expect that  Type-II bursts, which arise due to the shock acceleration
in the solar magnetosphere will be produced in the SGRs, since the plasma
density there is extremely small, so that except in the reconnection 
region, the magnetosphere is well-described by the force-free 
approximation, which does not allow the existence of shocks.

\section{Expected properties of SGR radio flares}

\subsection{Temporal behavior}

Energy release in reconnection appears to be a
 non-stationary transient phenomenon
resulting, presumably, from the spatially  fragmented structure.
The temporal behavior of Solar flares has several time-scales, associated
with different spatial scales of the reconnecting structures.
Similarly, the radio emission is expected to be  
non-stationary and  multi-times scaled,
 keeping the memory of the energy release history.

In reconnection, the 
 shortest time scale is related to the \Alfven crossing time of the
magnetic structures of length $L$: $\tau_r \sim L/v_A$ 
(for the Sun
this is $\sim 1$ sec).
The scale $L$ corresponds to the length of the reconnecting arc, which
for the SGRs may be as small as a fracturing of radius and as large as the
light cylinder radius. For flares occurring close to the surface we may
assume $L\sim R_{NS}$. The  \Alfven  velocity $v_A $ equals the speed of light
in the force-free magnetosphere. The observed rise time of the SGR X-ray
flares, $\leq 10$  msec, is consistent with being related to the
\Alfven time scale. For the observed bursts  the rise time
is limited by the intensity of the burst -
weaker bursts are expected to have shorter rise times (G\"og\"us 2002).
The shortest rise time is expected to be of the order of light  travel
time across the neutron star - tens of microseconds.
This time scale also gives the duration of shortest spikes in the
burst structure.  Radio bursts should have similar rise times, with a 
possible time-delay to allow for the plasma instabilities to develop
after the main X-ray burst.
 The overall duration of the burst depends on 
global structure of the reconnection region -- the
reconnection at one point may trigger reconnection at other points.

Radio emission should be more intermittent than
the X-ray emission, reflecting the fact the its
intensity depends both on the production rate, 
monitored well by the X-ray flux, and often subtle conditions 
for the development of kinetic instabilities (e.g. the requirement of beam
velocity to be larger than the thermal velocity  of the plasma
particles).

\subsection{Spectra}

Thompson et al. (2002)  discussed the properties of the strongly
twisted magnetosphere of the SGRs. Qualitatively, the maximum current
that the  magnetosphere can support corresponds to the toroidal field reaching
in strength
approximately  a poloidal field $B_\phi \leq B_p$. 
Below  we assume that such strong
  currents are indeed flowing in the SGR
magnetosphere (the strength of the current may be inferred from the
persistent luminosity of SGRs, see Thompson et al. 2002, eq. (34)).
 The velocity of the 
charge carriers  is weakly relativistic, $v\simeq c$. From the
induction equation we then find the current 
\be
j  \sim e n v \sim  c B / (4 \pi R )
\ee
and the plasma density $n$  and the plasma frequency $\om_p=\sqrt{ 4  \pi e^2 n /m} $:
\be
n \sim B/(  4  \pi e R), \, 
\om_p^2 \sim \om_B c /R 
\ee
Numerically, 
for $B_{NS} =  10^{14}$ G, $\om_B = \om_{B_{NS}} \tilde{r}^{-3}$, where
$\tilde{r} =  R/R_{NS}$, $ \om_{B_{NS}} = 2 \times 10^{21}$ rad/sec,
\be
\om_p = { \sqrt{  \om_{B_{NS}} c/R_{NS}}  \over  \tilde{r}^{2}} =
 7 \times 10^{10} (\tilde{r}/10)^{-2} {\rm rad/sec}
\label{w}
\ee
The self-similar model of Thompson et al. (2002) predicts that most of the 
non-potential energy of the magnetosphere is concentrated near the
stellar surface, at $R \leq 10 R_{NS}$.  Eq. (\ref{w}) may explain why
the radio emission from SGRs has not been detected yet and suggests 
the strategy for further searches. If the  coherent  radio emission is 
 generated near the stellar surface 
and is   associated with the local plasma frequency $\om_p$, then,
\footnote{Generation of coherent emission  at the plasma frequency
near th stellar surface is , in fact, {\it not} how we believe the
pulsar radio emission is produced, see below.} 
from eq. (\ref{w}), we may expect that the coherent radio emission should 
be generated at high frequencies, $\nu \geq 1 GHz$;
 below that  the plasma frequency is above the
observed frequency, so that plasma waves cannot propagate.

The 
radio emission of SGRs is expected to be  qualitatively different from the
normal radio pulsar emission. In  conventional radio pulsars the
presence of the primary beam with 
super-relativistic Lorentz factors is imperative
for the generation of radio  emission. In SGRs this primary beam may not 
be
created since the Goldreich-Julian density is much smaller than the density
of the currents required to support the twisted magnetic field.
If a large charge density is indeed generated on the open field lines, 
the particle accelerator, operating in the rotationally powered pulsars,
may be swamped and no pulsar-type radio emission is generated. 
This may be another reason why radio emission has not yet been detected from 
SGRs.

The radio emission of SGRs during bursting activity will
resemble  the solar radio Type-III bursts.
In solar Type-III bursts the energy is consecutively converted
from the   magnetic energy
into fast particles, then into  electrostatic 
plasma waves and finally  into escaping  electromagnetic waves.
The frequency of the generated 
EM waves is  the double of the plasma frequency
$\omega \sim 2 \om_p$.
Thus, one expects a {\it  narrow-band emission} $\Delta \om /\om \leq 1$.
The growth rate of Langmuir instability
\be
\Gamma \sim (n_b / n)^{1/3} \om_p \leq \om_p
\ee
($n_b$ is the beam density)
is indeed much higher than the dynamical time
\be
\Gamma /( c/R) \sim \sqrt{ \om_B /(c/R)} \gg 1
\ee
Thus, the plasma instability has enough time to develop.

A distinct feature   of the type-III burst is the drift of the
central frequency due to the spatial propagation of the
 emitting beam in the
inhomogeneous plasma.
Since the velocities of the emitting electrons are likely to be
 weakly relativistic,
the resulting emission may not be  narrow-band, as the electrons
propagate in the inhomogeneous plasma.
Still, one may expect the  
 frequency drift of the peak of radio emission,
 characteristic 
of  Type-III bursts.
 Since the plasma density in the SGR magnetosphere is
 $\om_p \sim \tilde{r}^2$, then if fast electrons
propagate with $v \sim c$, then the central frequency will move as
$
\om_{max} \sim t^{\pm 2}
$
taking into account the possibility of upward and downward movement.
The multi-polar structure of the magnetosphere may change this simple
dependence. 

It may also be possible to observe the U-type subclass of the Type-III bursts:
in this  subclass the central frequency first decreases and then
starts to increase as emitting electrons move along the closed field lines,
reach the maximum height above the stellar surface (at this point the
density is minimal, so is the frequency of emission) and then  return 
to the stellar surface.  

\subsection{Expected Flux}

The radio brightness of SGR bursts may be estimated using the 
energy partitioning in the solar flares, where 
the energy release in radio is typically $10^{-4}$ of the
energy released in hard X-rays
\footnote{The direct application of the  radio efficiency
of the Solar flares to magnetars is naturally only a guess.}
(plus an approximately
 similar amount of energy is released
in bulk motion, thermal heating and Cosmic rays). 
Since the X-ray luminosity of flares is $\sim 10^{36}-  10^{39}$ erg/sec,
the expected  radio  luminosity 
 $\sim 10^{33}-  10^{35}$ erg/sec, which at a distance $\sim 10 $ kpc
and the observing frequency $\sim 1 GHz$ will produce a flux of
$\sim 1 - 1000 Jy$, which 
 may be easily
detectable.

\section{Discussion}

We encourage radio  observations of SGRs and AXPs during their active phase
at high frequencies $\geq 1 GHz$.    
This requires catching a burst in simultaneous radio and X-ray 
observations.
\footnote{We would like to stress  that we predict a coherent emission 
{\it during} the bursts, not the cyclotron emission from the 
plerionic nebular after the burst (Frail et al. 1999)}
During its active phase
SGR 1900+14 produces bursts every $\sim 50$ seconds, emitting 
$\sim 10^{38} $ ergs/sec burst every $\sim 10$ minutes (G\"og\"us 2002).
 The radio flux
from such a burst $\sim 10$ Jy can be easily detected. Though
larger flares are less likely ($dN/dE \sim E^{-1.66}$) a flare 10 times stronger
is still observed once per hour. 
The search should be done in  a pulsar  mode with fast timing. 
Initial detection will naturally require a search in the DM space;
the frequency drift of the emission may complicate the DM search.
A strong correlation with the X-ray burst may provide an additional 
help in detecting radio  bursts, especially after the first one is seen and
time delay between the X-rays and radio due to the ISM propagation is measured.

Persistent radio  emission from SGRs may also be observed, though the expected
fluxes $\sim 1-10 $ mJy (based on the
 same radio/X-ray luminosity ratio of $10^{-4}$
and the persistent X-ray emission from SGRs  $ \sim 10^{34}-10^{35}$ ergs/sec)
may be too faint.

A number of factors may preclude the detection of radio emission
from SGR burst:
(i) reconnection in SGRs may be qualitatively different from the
reconnection on the Sun; (ii) radio emission may be 
strongly absorbed (or scattered at $r<r_a$)  at the cyclotron resonance
 inside the SGR magnetosphere.
Incidentally, if we assume that  a $\sim 6$ sec pulsar strongly
scatters or absorbed radio waves inside the light cylinder, then one would
expect a sharp cut-off above $\sim 100$ MHz, consistent with the
claims of detection of  SGR  1900+14
frequencies and non-detection at higher frequencies (Shitov et al 2000, 
Gil et al. 1998); (iii) abandant pair production during the burst
may significanly increase the plasma density and the plasma frequency,
pushing the radio emission to higher frequencies.

Another 
possible  mechanism of radio emission generation --
due to the loss-cone instability and at the  anomalous cyclotron-Cherenkov
resonance -- are not likely to operate in SGRs.
In the super-strong  magnetic fields of SGRs  electrons lose their transverse
energy almost immediately. Thus, we don't expect any adiabaticly
trapped electrons to exists near the neutron star radius (the
adiabatic radius, where the cyclotron decay time becomes equal rational period
is $r_{a} \sim 5 \times 10^{3} R_{SN} \sim 0.15 R_{LC} $.
 \footnote{
Cyclotron transition time $\tau= c/(r_e \om_B^2)$; it equals the rotational
period    
at $ r_a =(P \om_{B_{SN}}^2 r_e /c)^{1/6} R_{NS} $}
So no loss-cone instability will develop. 
Since the difference of the refractive index of plasma from unity 
is negligible $n-1\sim { c / R \om_B} \sim 10^{-18}$, no 
 anomalous cyclotron - Cherenkov instability  (Lyutikov et al.  1999)
will develop either.
 We can also neglect the (frequency-independent)
 dispersion inside the
magnetosphere.

\begin{acknowledgements} 
I would like to thank V. Kaspi, M. Roberts, F. Gavriil, C. Thompson, 
P. Woods, C. Kouveliotou and E. G\"og\"us 
 for interest in this work and Alissa Nedossekina for comments on the
manuscript. This research
was partly done at the  USRA/NSSTC; I am  greatful for their hospitality.
\end{acknowledgements}

\end{document}